\documentclass[journal]{IEEEtran}
\usepackage[utf8]{inputenc}
\usepackage{epsfig}    
\usepackage{amsfonts}
\usepackage{amssymb}
\usepackage{parskip}
\usepackage{amsmath}
\usepackage{comment}
\usepackage{multirow}
\usepackage{psfrag}
\usepackage[linesnumbered,ruled]{algorithm2e}
\usepackage{soul} 
\usepackage{float}
\usepackage{tikz}
\usetikzlibrary{arrows.meta, decorations.markings}
\usetikzlibrary{automata}
\usepackage{lipsum}  
\usepackage{subcaption}
\usepackage{diagbox}
\usepackage{url}
\usepackage{hyperref}
\usepackage{xcolor}

\usepackage{stfloats}
\usepackage{cite}
\usepackage[normalem]{ulem}

\newcommand{\nnp}[0]{\text{95th percentile }}

\definecolor{boristext}{rgb}{0.22, 0.44, 0.88}
\definecolor{boriscomments}{rgb}{0.88, 0.04, 0.04}
\definecolor{boristochange}{rgb}{0.2, 0.8, 0.8}

\begin{document}
\title{An Experimental Study of Latency for\\IEEE 802.11be Multi-link Operation}
\author{
\IEEEauthorblockN{Marc Carrascosa$^{\star}$, Giovanni Geraci$^{\star}$, Edward Knightly$^{\flat}$, and Boris Bellalta$^{\star}$}\\ \vspace{0.3cm}
\normalsize\IEEEauthorblockA{$^{\star}$Dept. Information and Communication Technologies, Universitat Pompeu Fabra, Barcelona}\\
$^{\flat}$Dept. Electrical and Computer Engineering and Computer Science, Rice University, Houston, TX
\thanks{M. Carrascosa and B. Bellalta were supported by WINDMAL PGC2018-099959-B-I00 (MCIU/AEI/FEDER,UE) and Cisco.}
\thanks{G. Geraci was supported by MINECO's Project RTI2018-101040 and by a ``Ram\'{o}n y Cajal" Fellowship from the Spanish State Research Agency.}
}

\maketitle


\begin{abstract}
Will Multi-Link Operation (MLO) be able to improve the latency of Wi-Fi networks? MLO is one of the most disruptive MAC-layer techniques included in the IEEE 802.11be amendment. It allows a device to use multiple radios simultaneously and in a coordinated way, providing a new framework to improve the WLAN throughput and latency. In this paper, we investigate the potential latency benefits of MLO by using a large dataset containing 5 GHz spectrum occupancy measurements.
Experimental results show that when the channels are symmetrically occupied,  MLO can improve latency by one order of magnitude. In contrast, in asymmetrically occupied channels, MLO can sometimes be detrimental and increase latency. To address this case, we introduce  Opportunistic Simultaneous Transmit and Receive (STR+) channel access and study its benefits.
\end{abstract}


\section{Introduction}

The importance of wireless connectivity in a globalized society is unquestionable, and forced lockdowns reminded us how dependable Wi-Fi is. We resorted to Wi-Fi to be in touch with our loved ones, to make online purchases, and to get work done and keep the economy afloat. In a post-pandemic world, Wi-Fi technologies will be vital for accessing fair and remote-friendly education, medical care, and business opportunities in the unlicensed spectrum. There will be nearly 628 million public Wi-Fi hotspots by 2023 \cite{Cisco2020}, one out of ten equipped with Wi-Fi 6 based on the IEEE 802.11ax amendment \cite{2018IeeeStandardAxDraft}.

As the popularity of Wi-Fi grows, so does the demand for augmented data rates, higher reliability, and lower latency, driving the development of a new Wi-Fi 7 generation based on the IEEE 802.11be Extremely High Throughput (EHT) specification \cite{draft11be, garcia2021ieee, khorov2020current, hoefel2020ieee, yang2020survey}. Despite its name, Wi-Fi 7 will be chasing much more than peak throughput. Indeed, the 802.11be Task Group acknowledges the need for lower delays to enable delay-sensitive networking use cases, including augmented and virtual reality, cloud computing, and cross-factory floor communications in next-generation enterprises \cite{8847238,bellalta2020low,adame2021time,carrascosa2020cloud,9152055}.

In a quest for lower delays, one of the most disruptive features being proposed for 802.11be is Multi-Link Operation (MLO) \cite{levitsky2020study,yang2019ap,DBLP:journals/corr/abs-2105-10199}. In MLO, devices can make simultaneous use of different channels or bands, potentially allowing delay-sensitive traffic to be transmitted through multiple links to ensure its timely reception. With its standardization process being consolidated, and prompted by the increasing interest from the research community \cite{song2021performance,9500256,9550829,9557495}, a fundamental question arises as to whether and to what extent MLO can reduce Wi-Fi latency in real-world scenarios.

In this paper, capitalizing on over-the-air measurements of spectrum occupancy for the entire 5~GHz band recently collected \cite{barrachina2020wi,barrachina2021wi} and freely available in open source\footnote{WACA dataset: \url{https://github.com/sergiobarra/WACA_WiFiAnalyzer}.}, we experimentally investigate the latency\footnote{The terms latency and delay are used interchangeably throughout the paper.} performance of 802.11be MLO. Atop these traces, which include scenarios with high access point (AP) density and crowded environments and span multiple hours, we develop an emulation tool that fuses a Wi-Fi MLO state machine with the high-resolution spectrum measurements. Besides legacy Wi-Fi Single-Link Operation (SLO), we study the two MLO channel access modes currently under consideration by the IEEE 802.11be Task Group \cite{draft11be}: (i) MLO-STR, where two radio interfaces are operated independently, and (ii) MLO-NSTR, where one interface acts as primary and the other as secondary. Our main contributions can be summarized as follows: 
\begin{itemize}
    \item We show that when using two links with statistically symmetrical occupancy, MLO reduces average and 95th percentile latency by up to 78\% with respect to SLO by availing of a second radio interface.
    \item In contrast, we surprisingly discover that when using two links with asymmetrical occupancy, MLO-STR can sometimes worsen the latency performance with respect to SLO. In the worst case, we observe an increase of up to 112\% in terms of \nnp latency. 
    \item To overcome the aforementioned issue, we propose an alternative implementation of STR, denoted MLO-STR+. By running in parallel as many backoff instances as interfaces, MLO-STR+ allocates each packet to the interface whose backoff expires first. This way, STR+ guarantees same delay as or lower than SLO, with reductions of up to 70\% in the best observed case. 
\end{itemize}

\section{Multi-radio Multi-link Operation}\label{multilinkbackground}

\begin{figure*}[ht]
    \centering
    \includegraphics[width= 0.75\textwidth]{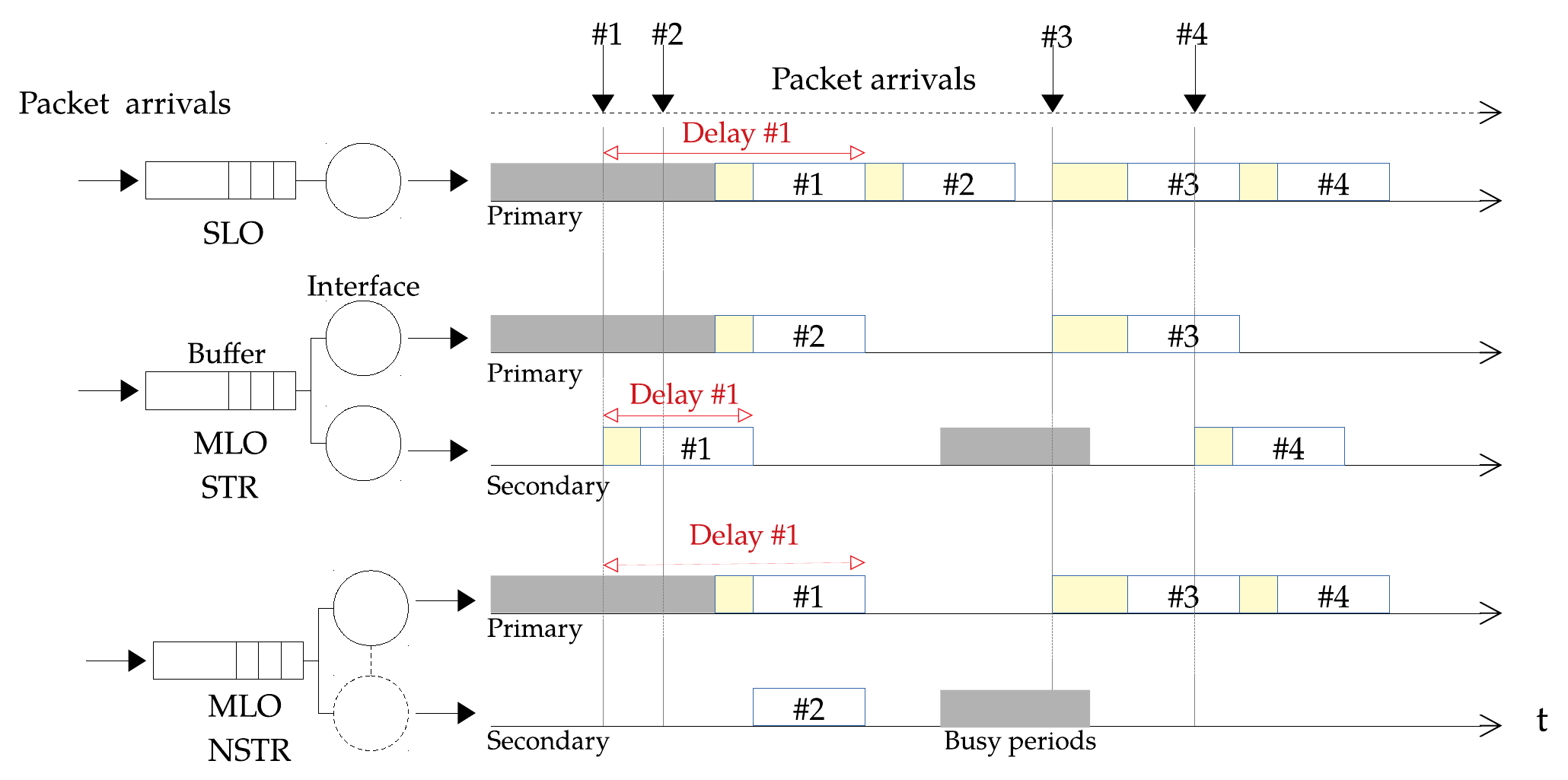}
    \caption{Illustration of SLO, MLO-STR, and MLO-NSTR operations. Grey, yellow, and white bars respectively indicate occupied channels, random backoffs, and packet transmissions. Packet transmissions include both the data part and the corresponding ACK, as well as DIFS and SIFS inter-frame spaces.}
    \label{ft3}
\end{figure*}

IEEE 802.11be considers two main channel access methods to support Multi-link Operation: Simultaneous Transmit and Receive (MLO-STR), and Non-simultaneous Transmit and Receive (MLO-NSTR) \cite{draft11be}. We introduce them in the following, from the perspective of an AP equipped with two radio interfaces and thus able to operate on two different channels simultaneously:

\begin{itemize}
    \item \textbf{MLO-STR:} The two radio interfaces operate independently and asynchronously, and a packet waiting for transmission is allocated to a radio interface as soon as the latter becomes available. If both radio interfaces are available, the packet is randomly allocated to either. Once an interface is allocated a packet, it starts channel contention by initializing a backoff instance.
    \item \textbf{MLO-NSTR:} One interface acts as primary, and the other as secondary. When there are packets waiting for transmission, the primary interface undergoes contention to access the channel through a backoff counter. Once the backoff counter reaches zero, packets are sent through both interfaces if the secondary one has been idle for at least a PIFS interval. Otherwise, only the primary interface is used to transmit.
\end{itemize}

Besides the MLO modes, IEEE 802.11be also considers the conventional Single-link Operation, where an AP is equipped with only one radio interface.

Figure~\ref{ft3} exemplifies SLO, MLO-STR, and MLO-NSTR operations. SLO follows default Wi-Fi operations, where packets are sequentially transmitted. In the case of MLO-STR, arriving packets are allocated to whichever interface becomes available first. This results in a significant delay reduction for packets \#1, \#2 and \#4. In the case of MLO-NSTR, the secondary channel's dependence on the primary sometimes prevents efficiently using the two radio interfaces. As a result, and unlike MLO-STR, the delay for packets \#1 and \#4 cannot be reduced with respect to SLO.


\section{Experimental Setup}\label{evaluationmethodology}

\begin{figure}
    \centering
    \includegraphics[width= 0.9\columnwidth]{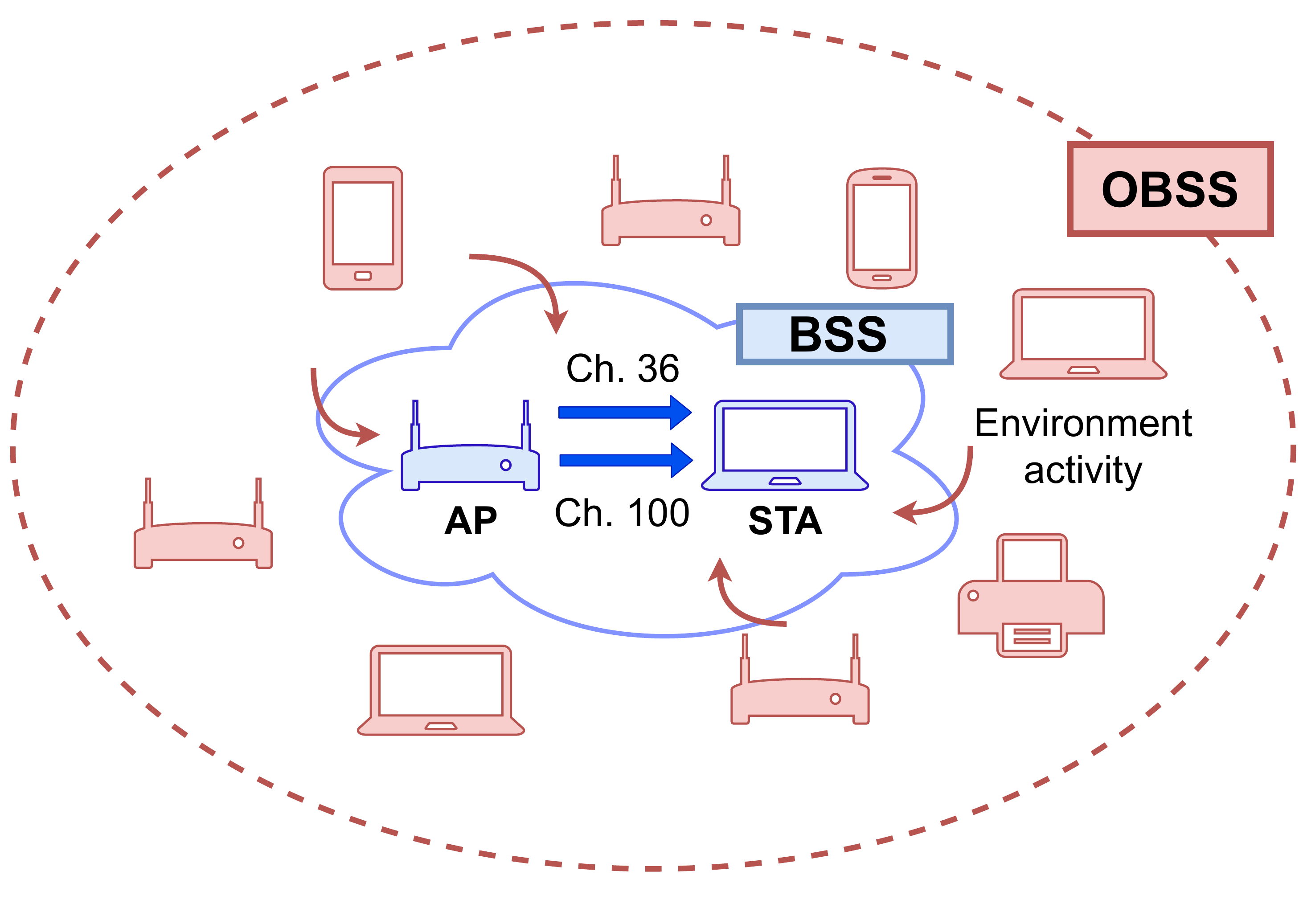}
    \caption{Scenario considered. The WACA dataset is used to characterize the environment activity (red) observed by the target BSS (blue) on channels 36 and 100 in the 5 GHz band. }
    \label{fig:system_model}
\end{figure}

In this work, we consider a target WLAN Basic Service Set (BSS) consisting of one AP and one station (STA), both equipped with two Wi-Fi interfaces each operating in the 5 GHz band on channels 36 and 100, respectively denoted \emph{primary} and \emph{secondary}. On these channels, the target BSS observes the environment activity, i.e., the transmissions generated by Orthogonal Basic Service Sets (OBSS). The BSS and OBSS under consideration are illustrated in Figure~\ref{fig:system_model} in blue and red, respectively. For this setup, we consider the three modes of operation described in Section~\ref{multilinkbackground}, namely: (i) SLO, where only the primary channel interface is available; (ii) MLO-STR, where both interfaces are available and work independently; and (iii) MLO-NSTR, where both interfaces are available but usage of the secondary channel is conditioned on the primary also being unoccupied.

For the above scenario, we consider downlink traffic, i.e., from the AP to the STA. We assume packet arrivals to follow a Poisson process, and transmitted packets to have a constant size of $L=12000$ bits. Table~\ref{teVar} summarizes the main parameters used in the Wi-Fi state machine. 

\subsection{WACA Dataset}

In order to evaluate the latency of 802.11be MLO in a real-world setting, we employ the WACA dataset, containing over-the-air measurements of the 5 GHz band occupancy that we have recently collected and made publicly available. This  dataset was obtained by conducting extensive measurement campaigns on different days and in multiple locations, including a sold-out football stadium (F.~C. Barcelona's Camp Nou). In this paper, we employ the football stadium measurements since they range from completely idle to fully occupied channels. In the dataset, spectrum samples consist of  1~s of consecutive, 10~$\mu$s receive signal strength indicator (RSSI) measurements. We refer the reader to \cite{barrachina2020wi,barrachina2021wi} for further details on the dataset. Compared to \cite{barrachina2020wi,barrachina2021wi}, in this work we have implemented a wholly new Wi-Fi state machine, capable of (i) fully characterizing the temporal dynamics of the system under finite traffic loads, i.e., non-full buffer conditions, and (ii) supporting multiple Wi-Fi interfaces and packet buffers.

\subsection{Trace-based Simulations Methodology}

In order to study the effect of channel occupancy on latency, we partition the available traces in our dataset for both primary and secondary channels into different average channel occupancy regimes: \{10\%, 20\%, \ldots, 90\%\}. Then, we run each simulation as follows:
\begin{enumerate}
    \item We select the occupancy regime of interest for the primary and secondary channels, e.g., 10\% and 40\%, respectively;
    \item We combine uniformly at random one spectrum sample each for the primary and secondary channels;
    \item For each spectrum sample pair and given a particular traffic load of interest, we compute the packet arrival times at the AP;
    \item We execute the Wi-Fi state machine for SLO, MLO-STR, and MLO-NSTR over each spectrum sample, considering for each channel access mode the same packet arrival times. We consider the same \emph{hinder interaction model} as in \cite{barrachina2021wi}.
    
    \item We store the individual delay experienced by each packet over all spectrum samples.
\end{enumerate}

For a fair comparison between SLO and MLO, we guarantee that all results are obtained in non-saturation conditions and thus we discard any simulations where less than 95\% of all the transmitted packets are received.

\begin{table}[b]\centering
	\begin{small}
    \begin{tabular}{|p{4cm}|c|c|}
  		\hline  
		\textbf{Name}  &   \textbf{Variable}& \textbf{Value}\\ \hline
		Legacy preamble  & $T_{\text{PHY-legacy}}$  & $20~\mu s$\\ \hline
		HE single-user preamble & $T_{\text{PHY-HE-SU}}$ & $52~\mu s$\\ \hline
		OFDM symbol duration& $\sigma$ & $16~\mu s$\\ \hline
		OFDM legacy symbol dur. & $\sigma_{\text{Legacy}}$ & $4~\mu s$\\ \hline
        Short InterFrame Space & SIFS & $16~\mu s$\\ \hline
        DCF InterFrame Space & DIFS & $30~\mu s$\\ \hline
        Slot time & $T_0$ & $10~\mu s$\\ \hline
		Service field   &  $L_{\text{SF}}$ & 32~bits\\ \hline
		MAC header   & $L_{\text{MH}}$ & 272~bits\\ \hline
		Tail bits & $L_{\text{TB}}$& 6~bits\\  \hline
		ACK bits & $L_{\text{ACK}}$ & 112~bits \\ \hline
		Frame size &  $L$ & 12000~bits\\\hline
	\end{tabular}
	\caption{Notation and Wi-Fi state machine parameters.} 
	\label{teVar}
	\end{small}
\end{table}




\section{Delay Performance}

This section investigates the delay performance of both SLO and MLO modes for different combinations of channel occupancies and traffic loads. 

To evaluate the gains of MLO in terms of delay, we consider the same traffic load for both SLO and MLO modes. In addition, the primary channel considered in the MLO mode is the same channel used in SLO. Therefore, we could expect that adding another channel in MLO mode, regardless of its occupancy, should yield lower delays.

\subsection{Symmetrically Occupied Channels}

\begin{figure*}[ht]
\centering
\begin{subfigure}[b]{0.32\textwidth}
    \includegraphics[width = \textwidth]{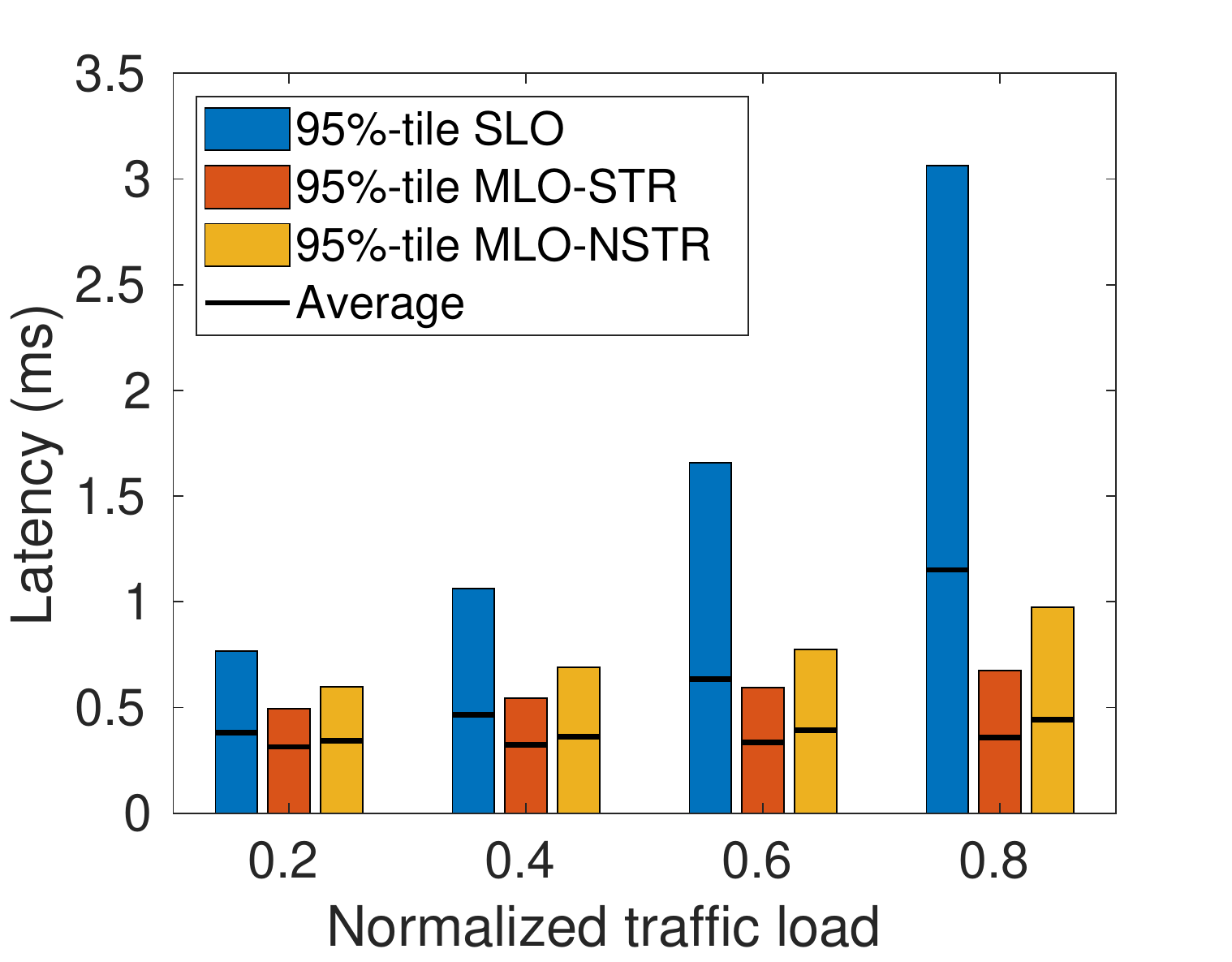}
    \caption{ 10\% occupancy on both channels   }
    \label{sym1}
\end{subfigure}
\begin{subfigure}[b]{0.32\textwidth}
    \includegraphics[width = \textwidth]{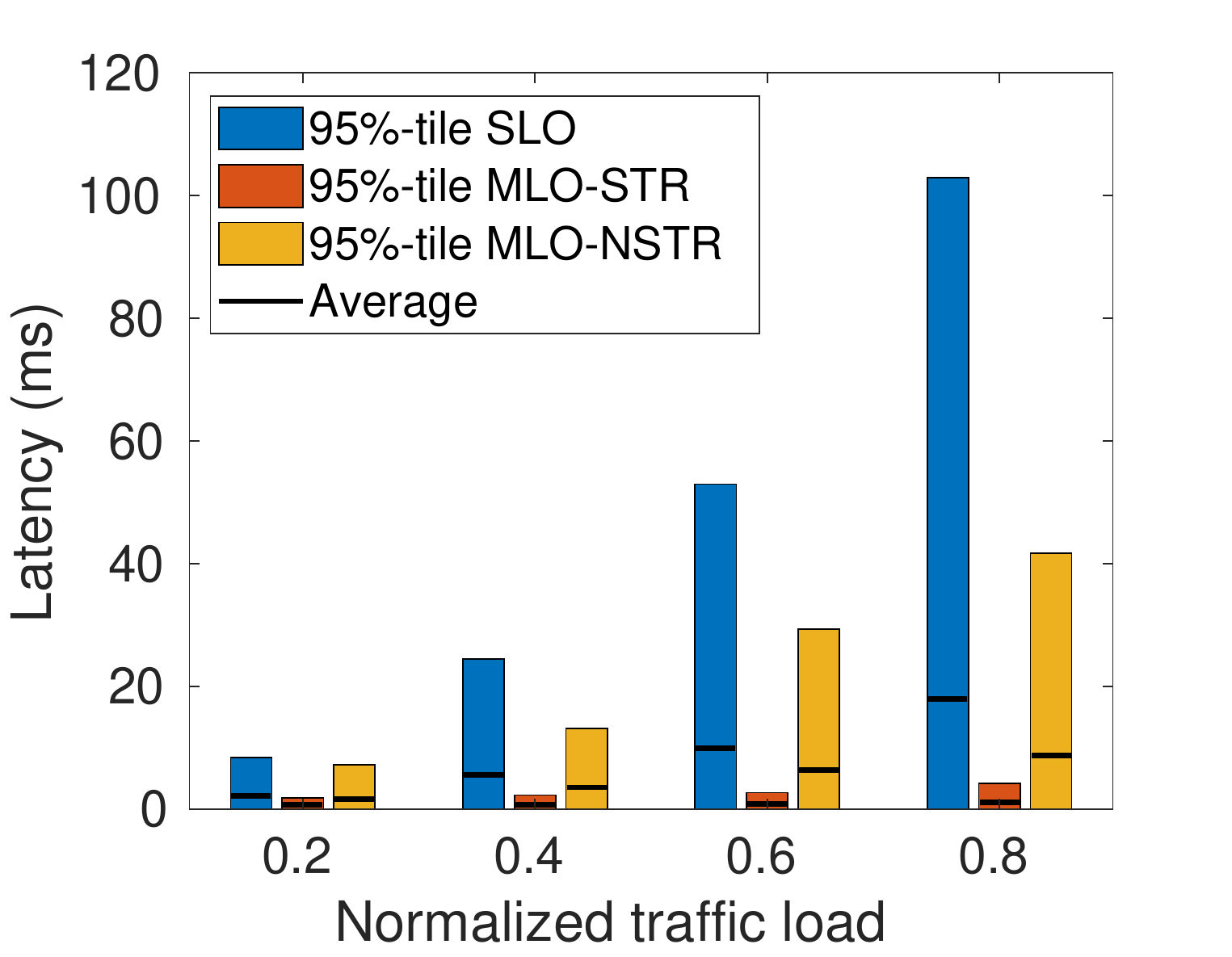}
    \caption{ 40\% occupancy on both channels}
    \label{sym4}
\end{subfigure}
\begin{subfigure}[b]{0.32\textwidth}
    \includegraphics[width = \textwidth]{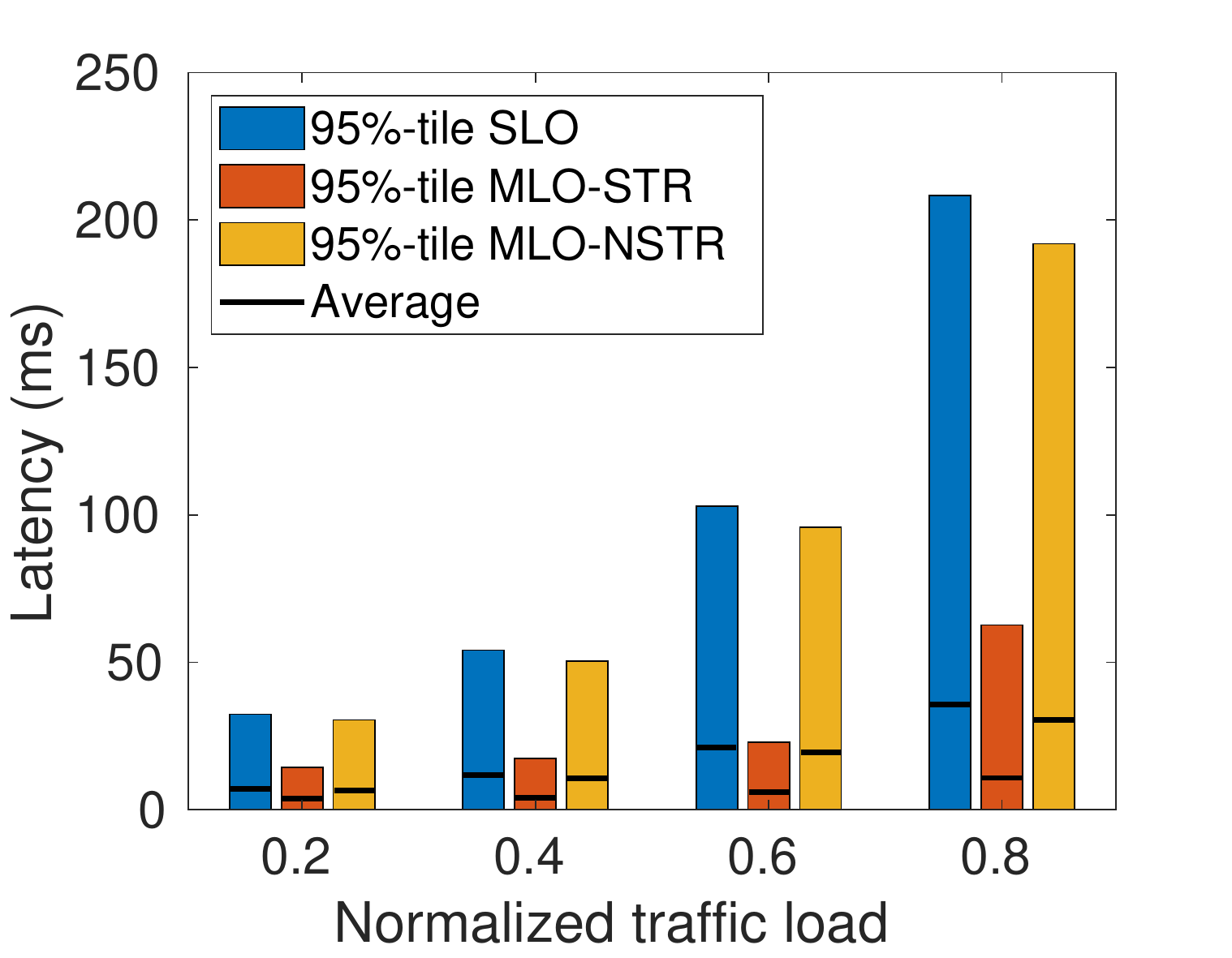}
    \caption{70\% occupancy on both channels}
    \label{sym7}
\end{subfigure}
\caption{Latency for symmetrically occupied channels vs. variable normalized traffic load.}
\label{syms1}
\end{figure*}

Here we study the case of \emph{symmetric} channel occupancies in which both MLO interfaces have channels with similar occupancy levels. In particular, we study the delay performance with pairs of channels in the ranges of 10\%, 40\%, and 70\% occupancy. In those cases, the average full-buffer throughput under SLO is 37, 22, and 6.8 Mbps, respectively. For these three scenarios (symmetric low, medium, and high occupancy), we feed the Wi-Fi state machine with Poisson traffic and vary the intensity as a fraction of this SLO average full-buffer throughput, namely {0.2, 0.4, 0.6, 0.8}.
 
Figure~\ref{syms1} shows the average and 95th percentile delay for all channel access modes and the different channel occupancies. First, we observe that when both channels have 10\% occupancy (Figure \ref{sym1}), the three schemes have strikingly different scaling with increasing traffic load as MLO delay does not increase at the same rate as SLO delay. For example, at 20\% traffic load, STR and NSTR offer a modest decrease in average delay compared to Single Link Operation of 17\% and 9\% respectively. In contrast, when the traffic load is 80\%, STR and NSTR reduce the average delay by 69\% and 62\%. This scaling is even more pronounced analyzing the 95th percentile of delay, in which MLO  achieves up to a 78\% delay reduction. Thus, for both average and 95th percentile delay, the benefits of MLO are increasingly pronounced under higher traffic load as in this case, there are often multiple packets in the buffer such that both interfaces can be used. Moreover, with a relatively low channel occupancy of 10\%, both channels are often available.  

Next, we consider the case that both channels have symmetrical medium (40\%) occupancy (Figure \ref{sym4}). Here, while SLO's \emph{average} delay increases only modestly with traffic (i.e., from 2 to 18~ms), the \nnp delay increases much more rapidly, exceeding 100~ms. In contrast, STR can yield a staggering order of magnitude reduction in \nnp delay compared to SLO. The reason is that STR avails usage of two channels and can access either or both of them. STR, therefore, realizes delay benefits compared to SLO unless both channels are occupied. 

Unfortunately, unlike STR, the benefits of NSTR over SLO are limited and mostly confined to how the average delay scales with traffic load. Indeed, NSTR can only gain access to the secondary channel if the primary channel is also idle, implying that the average delay is guaranteed to be lower than the average delay under SLO. However, the \nnp delays are triggered by long periods of occupancy of the primary channel, thus making any availability of the secondary channel during this time irrelevant. As a result, the \nnp delay under NSTR rapidly grows as the normalized traffic load increases. 

Lastly, when both channels have high (70\%) occupancy (Figure \ref{sym7}), STR again has the most favorable \nnp delay scaling with traffic load, providing substantial reductions as compared to both SLO and NSTR. Nonetheless, at such high channel occupancies, even STR has difficulty finding transmission opportunities on either channel, so both mean and \nnp delays are increasing. Additionally, NSTR provides negligible benefits in both average and \nnp delay compared to SLO.

\textit{Findings:} When both channel occupancies are symmetrically medium to high load, NSTR fails to provide significant \nnp delay benefits compared to SLO. The key reason is that NSTR is only able to realize a benefit compared to SLO if both channels are simultaneously unoccupied, an increasingly unlikely occurrence in this scenario. Fortunately,  STR yields significant \nnp latency benefits (compared to both SLO and NSTR) even in the challenging regime of increasing occupancies and traffic. This is because STR can utilize either available channel, and reduce the packets waiting time even if it cannot simultaneously utilize both available channels.

\subsection{Asymmetrically Occupied Channels}\label{wcd}

\begin{figure*}[ht]
\centering
\begin{subfigure}[b]{0.32\textwidth}
    \includegraphics[width = \textwidth]{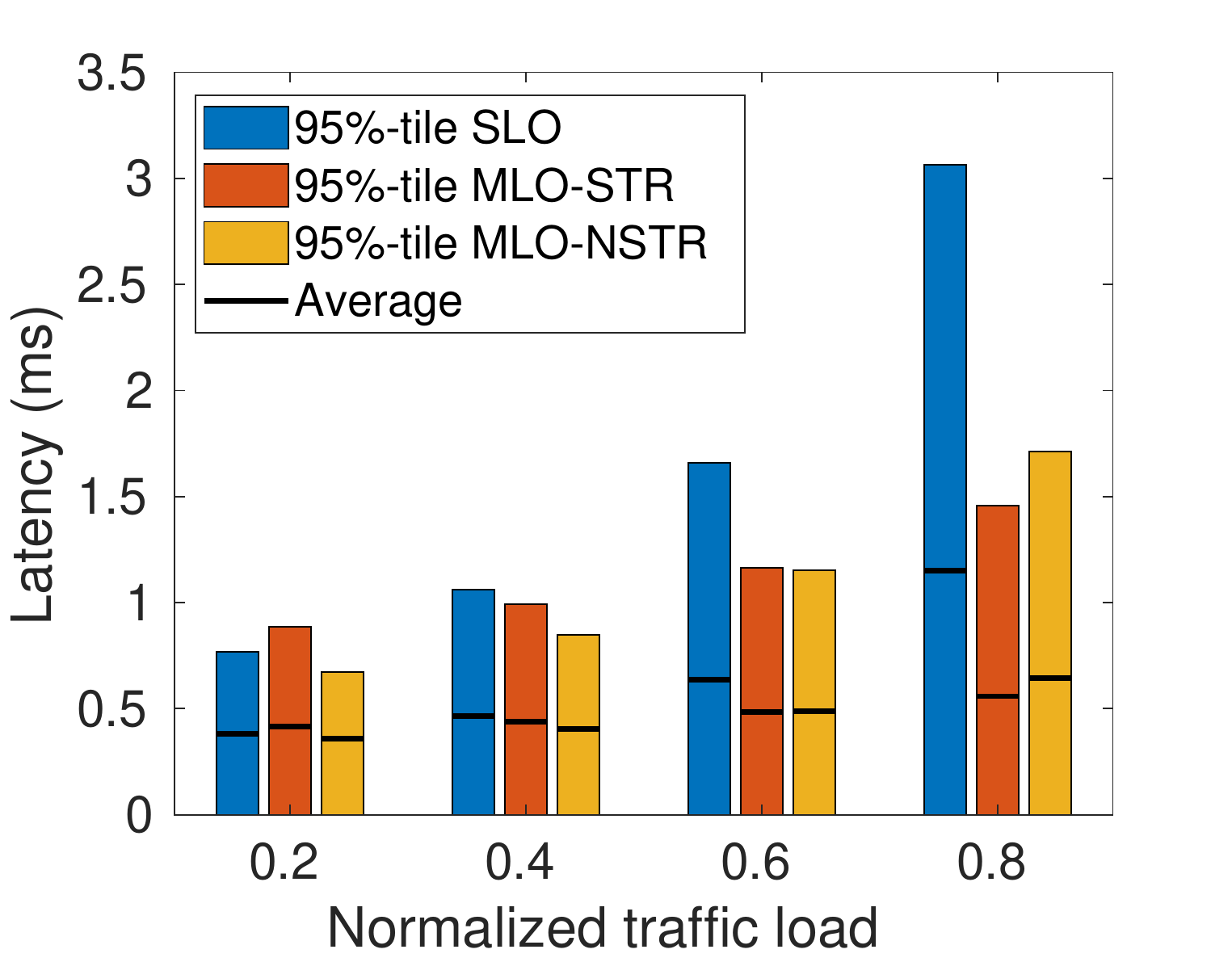}
    \caption{ Primary of 10\% and secondary of 40\% }
    \label{delay1n}
\end{subfigure}
\begin{subfigure}[b]{0.32\textwidth}
    \includegraphics[width = \textwidth]{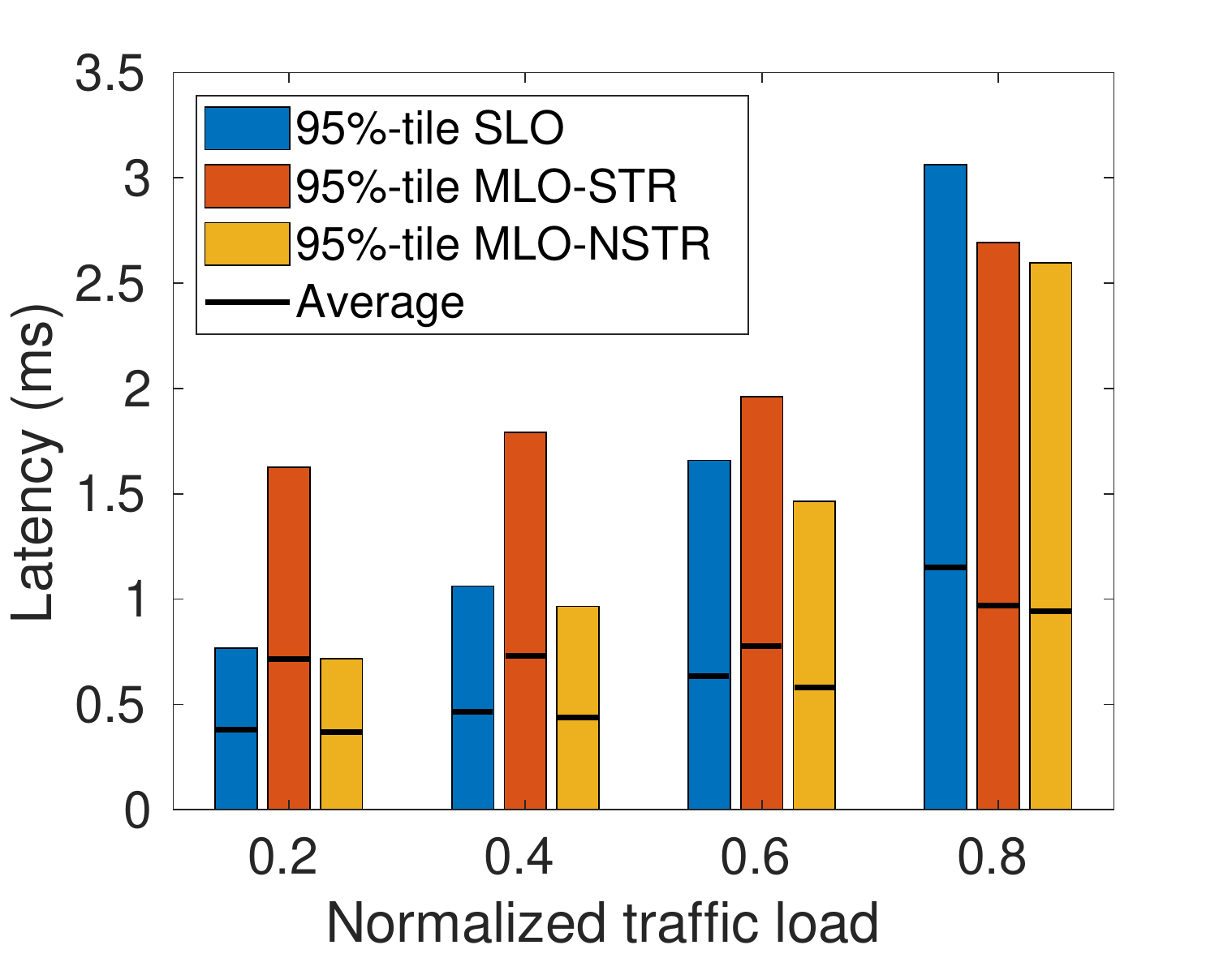}
    \caption{Primary of 10\% and secondary of 70\%}
    \label{delay3n}
\end{subfigure}
\begin{subfigure}[b]{0.32\textwidth}
    \includegraphics[width = \textwidth]{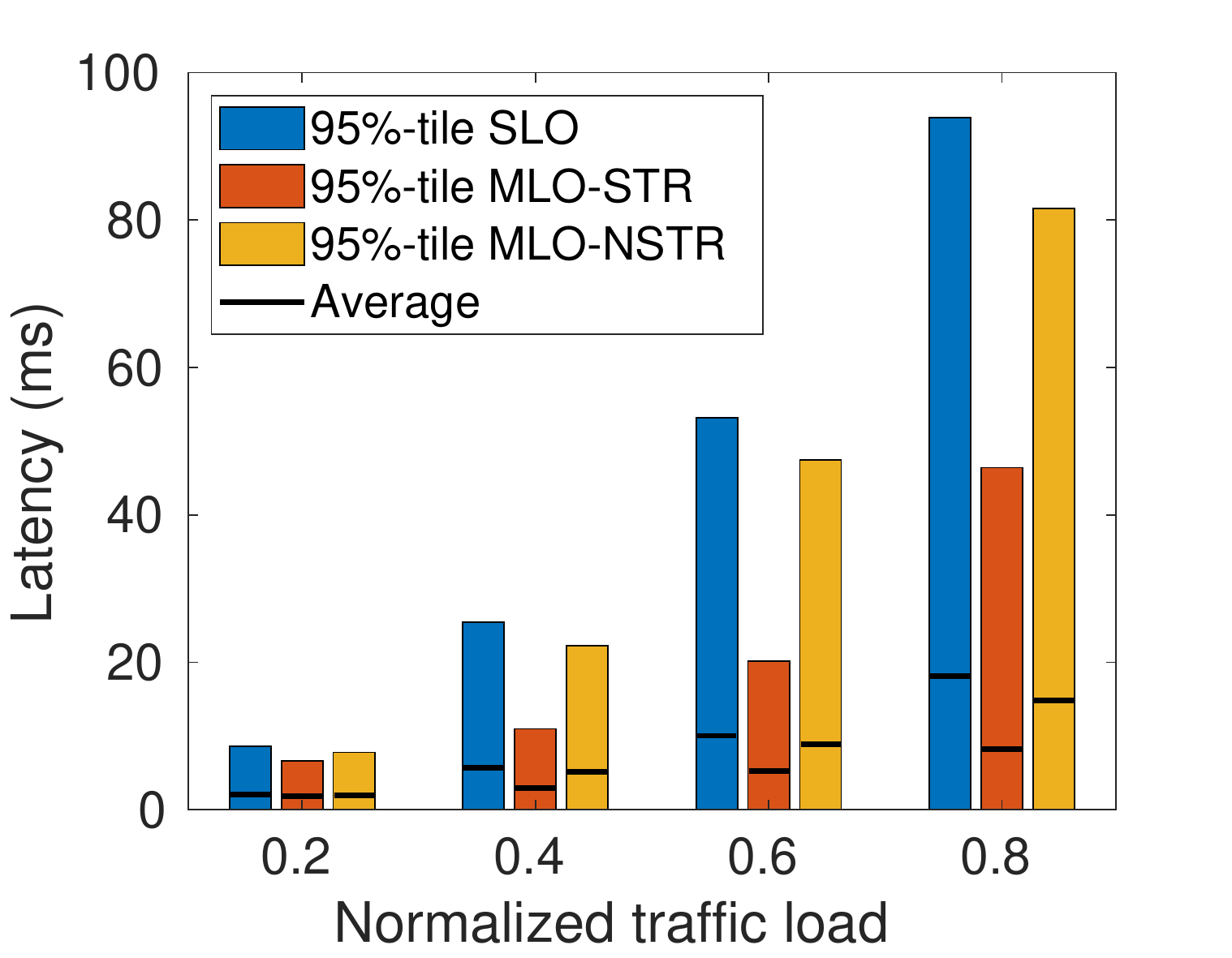}
    \caption{Primary of 40\% and secondary of 70\%}
    \label{delay5n}
\end{subfigure}
\caption{Latency for non-symmetrically occupied channels vs. variable normalized traffic load.}
\label{delsn}
\end{figure*}

Here, we employ the same normalized traffic loads from the previous section, but change the channel occupancy of our interfaces so that they lie in different ranges. Between the two channels, we always assume the primary to be the less occupied one. Note that the opposite case favors both MLO modes in this comparison: SLO would always incur a high delay, and both MLO modes would take advantage of a more idle secondary channel.

Figure~\ref{delay1n} depicts the case of a low (10\%) primary and medium (40\%) secondary channel occupancy. As expected, NSTR offers deterministically lower delays than SLO, with the highest benefits occurring under higher traffic loads. However, STR surprisingly incurs a higher average and \nnp delay than SLO for the lowest traffic load of 0.2. Indeed, STR starts contention by initializing the backoff counter as soon as a channel is detected to be idle. Unfortunately, such channel may be occupied before the backoff timer expires, thus pausing the backoff counter. If the backoff is paused too often (or for long intervals), the packet could incur even higher delays than it would have if the other channel---initially busy---had been selected.

In Figure~\ref{delay3n}, this effect is exacerbated due to the even higher occupancy of the secondary channel, as selecting an idle secondary channel incurs the risk of the latter being occupied before the backoff counter expires. When this occurs, the \nnp delay can be twice as high as that with SLO, albeit still confined to below 10~ms. However, STR average and \nnp delays grow at a lower rate than those of SLO as the traffic load increases. Indeed, STR can still take advantage of a secondary channel (even when highly occupied) to reduce congestion and curb the latency when it is caused not only by the channel occupancy patterns but also by the amount of traffic.

Finally, Figure \ref{delay5n} considers the more symmetrical case of primary and secondary channel occupancy of 40\% and 70\%, respectively. Similar to the prior case of Figures~\ref{sym4} and \ref{sym7}, STR scales well with the increasing traffic load, keeping the average delay below that of SLO and the decreasing the 95th percentile by up to a half. Compared to Figure~\ref{delay3n}, the primary channel occupancy has grown, leading to a faster increase in the SLO delay vs. traffic load. However, STR is capable of leveraging both links and thus achieves lower delays. 

\textit{Findings:} Channel occupancy is a crucial factor to account for when selecting a secondary channel in MLO mode. For STR, specifically, using a secondary channel that is much busier than the primary can lead to even higher delays than using SLO. This is owed to packets being suboptimally assigned to an interface before carrying out the backoff, with the latter likely to be interrupted on the busier channel. This effect is exacerbated when the difference between channel occupancies increases.

\subsection{Opportunistic MLO-STR}

We have shown that MLO-STR can lead to even higher delays than using SLO in the case of channels with different occupancies. To overcome this issue, we now propose an alternative implementation which we denote as \emph{Opportunistic MLO-STR} (MLO-STR+):
\begin{itemize}
    \item \textbf{MLO-STR+:} When both interfaces are idle, one backoff instance is started on each. Packet allocation is deferred until either backoff counter expires, and the first waiting packet is allocated to the interface whose counter expires first. This approach differs from MLO-STR, where a packet is assigned to a channel as soon as the latter is idle, without waiting for its backoff counter to expire.
\end{itemize}

The main advantage of the proposed STR+ lies in the fact that, if one channel becomes occupied during the backoff, a transmission opportunity may be found on the other, avoiding unnecessarily delaying the waiting packet. In practice, implementing MLO-STR+ only requires a minor firmware update on the current Wi-Fi state machine: the ability to control when an interface can initiate, pause, and complete the backoff countdown without actually being allocated a packet.

\begin{figure*}[ht]
\centering
\begin{subfigure}[b]{0.32\textwidth}
    \includegraphics[width = \textwidth]{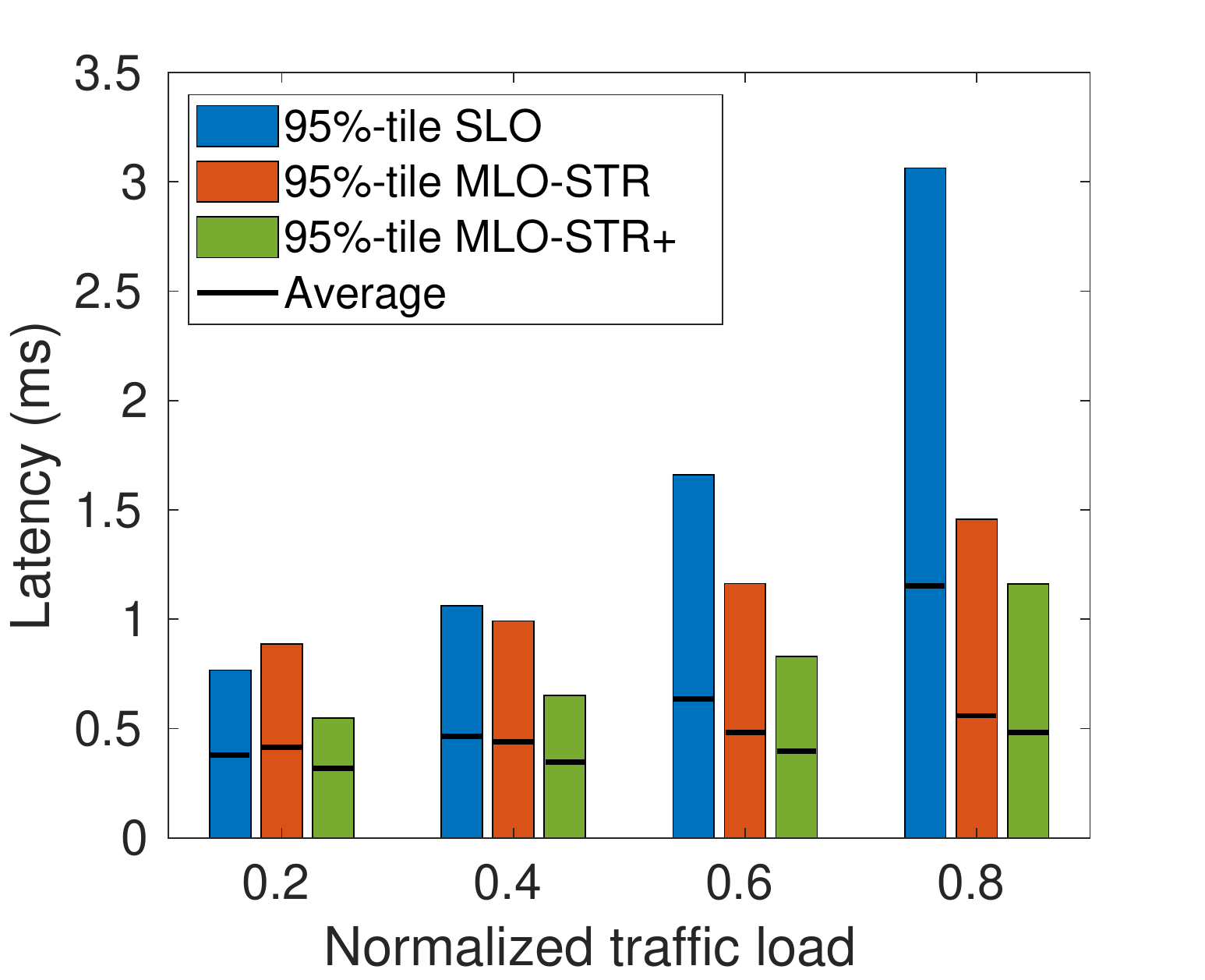}
    \caption{Primary of 10\% and secondary of 40\%  }
    \label{delay+}
\end{subfigure}
\begin{subfigure}[b]{0.32\textwidth}
    \includegraphics[width = \textwidth]{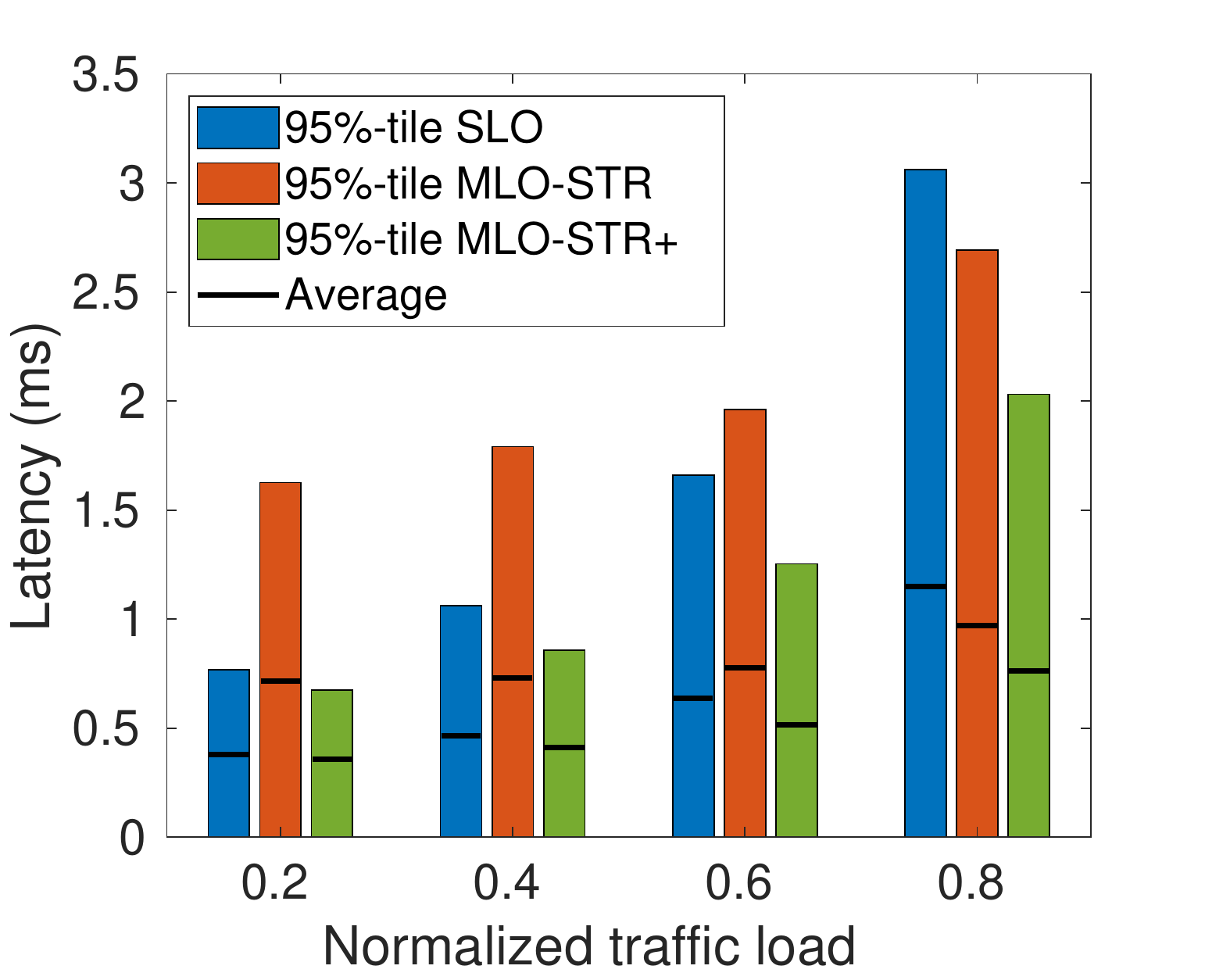}
    \caption{Primary of 10\% and secondary of 70\%}
    \label{nine}
\end{subfigure}
\begin{subfigure}[b]{0.32\textwidth}
    \includegraphics[width = \textwidth]{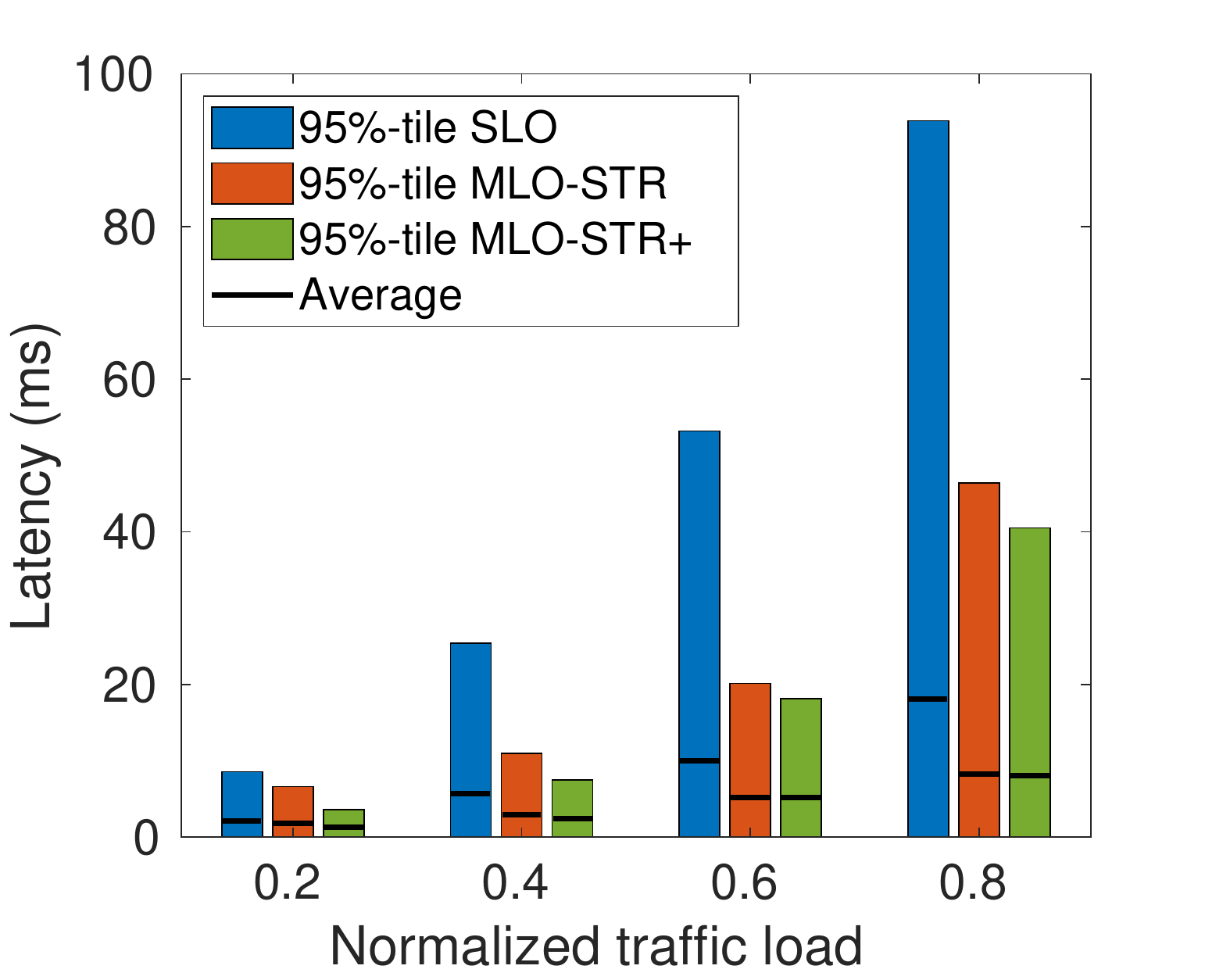}
    \caption{Primary of 40\% and secondary of 70\%}
    \label{delay3+}
\end{subfigure}
\caption{Latency for non-symmetrically occupied channels vs. variable normalized traffic load. MLN-STR vs MLN-STR+.}
\label{delsn+}
\end{figure*}

Figure~\ref{delsn+} shows the average and 95th percentile delay for the same cases studied in Figure \ref{delsn}. We still take SLO as the baseline and compare MLO-STR and MLO-STR+ modes. In Figure~\ref{delay+}, STR+ consistently outperforms STR and SLO in both average and 95th percentile delay, since packets are transmitted either at the same time as in SLO, or faster via the secondary interface.

In Figure~\ref{nine}, when the secondary channel has a 70\% occupancy, we encounter the worst scenario for STR. In this case, STR selects the secondary channel when it undergoes a short idle periods. However, since the latter are typically followed by longer intervals of occupancy, the backoff counter often remains frozen, leading to 95th percentile delays more than twice as high as those with SLO. This shortcoming is avoided altogether by the proposed STR+, assigning a packet to either interface only after ensuring that the corresponding backoff counter has expired.

Finally, Figure \ref{delay3+} depicts the case of 40\% and 70\% occupancy on the primary and secondary channels, respectively. As the former has increased, the SLO delay grows rapidly. STR already outperforms SLO in average and 95th percentile delay, and STR+ slightly reduces these values further.

\textit{Findings:} Multiple channels can be better exploited by running several backoff counters in parallel and allocating a waiting packet to whichever interface whose counter expires first. This proposed approach, which we denote MLO-STR+, gains more frequent transmission opportunities in the face of bursty spectrum occupancy, and guarantees a delay reduction with respect to SLO. 






\section{Conclusions}

In this paper, we provided an experimental study of latency for IEEE 802.11be MLO. Using the WACA dataset, which contains real-world channel occupancy measurements in the 5 GHz spectrum, we cast light upon the latency performance of two MLO channel access modes, namely (i) MLO-STR, where two radio interfaces are operated independently, and (ii) MLO-NSTR, where one interface acts as primary and the other as secondary. We showed that when both channels are on average equally occupied, both MLO modes can reduce the \nnp latency by nearly one order of magnitude as they avail of a second radio interface. In contrast, in asymmetrically occupied channels, we surprisingly found the use of MLO-STR to be detrimental and cause even higher latency values than SLO. To overcome this issue, we proposed an alternative implementation, denoted MLO-STR+. By running in parallel as many backoff instances as interfaces, MLO-STR+ allocates each packet to the interface whose backoff expires first, thus achieving a significantly lower latency.

\bibliographystyle{IEEEtran}

\bibliography{main.bbl}

\begin{thebibliography}{10}
\providecommand{\url}[1]{#1}
\csname url@samestyle\endcsname
\providecommand{\newblock}{\relax}
\providecommand{\bibinfo}[2]{#2}
\providecommand{\BIBentrySTDinterwordspacing}{\spaceskip=0pt\relax}
\providecommand{\BIBentryALTinterwordstretchfactor}{4}
\providecommand{\BIBentryALTinterwordspacing}{\spaceskip=\fontdimen2\font plus
\BIBentryALTinterwordstretchfactor\fontdimen3\font minus
  \fontdimen4\font\relax}
\providecommand{\BIBforeignlanguage}[2]{{%
\expandafter\ifx\csname l@#1\endcsname\relax
\typeout{** WARNING: IEEEtran.bst: No hyphenation pattern has been}%
\typeout{** loaded for the language `#1'. Using the pattern for}%
\typeout{** the default language instead.}%
\else
\language=\csname l@#1\endcsname
\fi
#2}}
\providecommand{\BIBdecl}{\relax}
\BIBdecl

\bibitem{Cisco2020}
{Cisco Annual Internet Report (2018--2023) White Paper}, Mar. 2020.

\bibitem{2018IeeeStandardAxDraft}
{IEEE 802.11}, ``{P802.11ax - IEEE Draft Standard for Information Technology --
  Telecommunications and Information Exchange Between Systems Local and
  Metropolitan Area Networks -- Specific Requirements Part 11: Wireless LAN
  Medium Access Control (MAC) and Physical Layer (PHY) Specifications Amendment
  Enhancements for High Efficiency WLAN},'' 2020.

\bibitem{draft11be}
``{IEEE P802.11be/D1.0 Draft Standard for Information technology—
  Telecommunications and information exchange between systems Local and
  metropolitan area networks— Specific requirements. Part 11: Wireless LAN
  Medium Access Control (MAC) and Physical Layer (PHY) Specifications.
  Amendment 8: Enhancements for extremely high throughput (EHT)},'' May 2021.

\bibitem{garcia2021ieee}
A.~Garcia-Rodriguez, D.~Lopez-Perez, L.~Galati-Giordano, and G.~Geraci, ``{IEEE
  802.11 be: Wi-Fi 7 Strikes Back},'' \emph{{IEEE Communications Magazine}},
  vol.~59, no.~4, pp. 102--108, 2021.

\bibitem{khorov2020current}
E.~Khorov, I.~Levitsky, and I.~F. Akyildiz, ``{Current status and directions of
  IEEE 802.11 be, the future Wi-Fi 7},'' \emph{{IEEE Access}}, vol.~8, pp.
  88\,664--88\,688, 2020.

\bibitem{hoefel2020ieee}
R.~P.~F. Hoefel, ``{IEEE 802.11 be: Throughput and Reliability Enhancements for
  Next Generation WI-FI Networks},'' in \emph{{2020 IEEE 31st Annual
  International Symposium on Personal, Indoor and Mobile Radio
  Communications}}.\hskip 1em plus 0.5em minus 0.4em\relax {IEEE}, 2020, pp.
  1--7.

\bibitem{yang2020survey}
M.~Yang and B.~Li, ``{Survey and perspective on extremely high throughput (EHT)
  WLAN—IEEE 802.11 be},'' \emph{Mobile Networks and Applications}, vol.~25,
  no.~5, pp. 1765--1780, 2020.

\bibitem{8847238}
D.~Lopez-Perez, A.~Garcia-Rodriguez, L.~Galati-Giordano, M.~Kasslin, and
  K.~Doppler, ``{IEEE 802.11be Extremely High Throughput: The Next Generation
  of Wi-Fi Technology Beyond 802.11ax},'' \emph{IEEE Communications Magazine},
  vol.~57, no.~9, pp. 113--119, 2019.

\bibitem{bellalta2020low}
B.~Bellalta, ``{On the Low-Latency Region of Best-Effort Links for
  Delay-Sensitive Streaming Traffic},'' \emph{{IEEE Communications Letters}},
  vol.~25, no.~3, pp. 970--974, 2020.

\bibitem{adame2021time}
T.~Adame, M.~Carrascosa-Zamacois, and B.~Bellalta, ``{Time-sensitive networking
  in IEEE 802.11 be: On the way to low-latency WiFi 7},'' \emph{Sensors},
  vol.~21, no.~15, p. 4954, 2021.

\bibitem{carrascosa2020cloud}
M.~Carrascosa and B.~Bellalta, ``{Cloud-gaming: Analysis of Google stadia
  traffic},'' \emph{arXiv preprint arXiv:2009.09786}, 2020.

\bibitem{9152055}
C.~Deng, X.~Fang, X.~Han, X.~Wang, L.~Yan, R.~He, Y.~Long, and Y.~Guo, ``{IEEE
  802.11be Wi-Fi 7: New Challenges and Opportunities},'' \emph{IEEE
  Communications Surveys Tutorials}, vol.~22, no.~4, pp. 2136--2166, 2020.

\bibitem{levitsky2020study}
I.~Levitsky, Y.~Okatev, and E.~Khorov, ``{Study on Simultaneous Transmission
  and Reception on Multiple Links in IEEE 802.11 be networks},'' in \emph{2020
  International Conference Engineering and Telecommunication (En\&T)}.\hskip
  1em plus 0.5em minus 0.4em\relax IEEE, 2020, pp. 1--4.

\bibitem{yang2019ap}
M.~Yang, B.~Li, Z.~Yan, and Y.~Yan, ``{AP Coordination and Full-duplex enabled
  Multi-band Operation for the Next Generation WLAN: IEEE 802.11 be (EHT)},''
  in \emph{2019 11th International Conference on Wireless Communications and
  Signal Processing (WCSP)}.\hskip 1em plus 0.5em minus 0.4em\relax IEEE, 2019,
  pp. 1--7.

\bibitem{DBLP:journals/corr/abs-2105-10199}
\BIBentryALTinterwordspacing
{\'{A}}.~L{\'{o}}pez{-}Ravent{\'{o}}s and B.~Bellalta, ``{IEEE} 802.11be
  multi-link operation: When the best could be to use only a single
  interface,'' \emph{CoRR}, vol. abs/2105.10199, 2021. [Online]. Available:
  \url{https://arxiv.org/abs/2105.10199}
\BIBentrySTDinterwordspacing

\bibitem{song2021performance}
T.~Song and T.~Kim, ``{Performance Analysis of Synchronous Multi-Radio
  Multi-Link MAC Protocols in IEEE 802.11 be Extremely High Throughput
  WLANs},'' \emph{Applied Sciences}, vol.~11, no.~1, p. 317, 2021.

\bibitem{9500256}
G.~Naik, D.~Ogbe, and J.-M.~J. Park, ``{Can Wi-Fi 7 Support Real-Time
  Applications? On the Impact of Multi Link Aggregation on Latency},'' in
  \emph{ICC 2021 - IEEE International Conference on Communications}, 2021, pp.
  1--6.

\bibitem{9550829}
H.~Park and C.~You, ``{Latency Impact for Massive Real-Time Applications on
  Multi Link Operation},'' in \emph{2021 IEEE Region 10 Symposium (TENSYMP)},
  2021, pp. 1--5.

\bibitem{9557495}
G.~Lacalle, I.~Val, O.~Seijo, M.~Mendicute, D.~Cavalcanti, and
  J.~Perez-Ramirez, ``{Analysis of Latency and Reliability Improvement with
  Multi-Link Operation over 802.11},'' in \emph{{2021 IEEE 19th International
  Conference on Industrial Informatics (INDIN)}}, 2021, pp. 1--7.

\bibitem{barrachina2020wi}
S.~Barrachina-Mu{\~n}oz, B.~Bellalta, and E.~Knightly, ``{Wi-Fi All-Channel
  Analyzer},'' in \emph{Proceedings of the 14th International Workshop on
  Wireless Network Testbeds, Experimental evaluation \& Characterization},
  2020, pp. 72--79.

\bibitem{barrachina2021wi}
S.~Barrachina-Mu{\~n}oz, B.~Bellalta, E.~W. Knightly \emph{et~al.}, ``{Wi-Fi
  Channel Bonding: An All-Channel System and Experimental Study From Urban
  Hotspots to a Sold-Out Stadium},'' \emph{IEEE/ACM Transactions on
  Networking}, 2021.

\end{thebibliography}

\end{document}